\documentclass{article}
\usepackage{graphicx}

\usepackage{amssymb}
\usepackage{setspace}

\begin{document}

\date{19, May 2009}
\title{Electrostatic and thermal fluctuations are the source of  magnetic fields in
unmagnetized inhomogeneous plasmas}
\author{Hamid Saleem\\National Centre for Physics (NCP),\\ Quaid-i-Azam University Campus,Islamabad,
 \\Pakistan. }

\maketitle

It is pointed out that electron thermal fluctuations can couple with
the ion acoustic mode in an inhomogeneous plasma to generate a low
frequency ion time scale electromagnetic wave. This electromagnetic
wave can become unstable if the temperature and density gradients
are parallel to each other which can be the case in laser-plasmas
similar to stellar cores. The comparisons of the present theoretical
model with the previous investigations are also presented. The final
result is applied to a classical laser induced plasma for
illustration.

\textbf{I. Introduction}\\ A low frequency electromagnetic wave
should be a normal mode of unmagnetized inhomogeneous plasmas as an
intrinsic source of magnetic fields. The ion acoustic wave (IAW) is
a very important mode because it exits in both unmagnetized and
magnetized plasmas. It is a low frequency wave and light electrons
are generally believed to act as inertia-less in its electrostatic
field.\\ In fact the electron inertia can play a vital role in
producing a low frequency electromagnetic wave. More than a decade
ago [1], it was proposed that an electromagnetic wave having
frequency near IAW is a normal mode of unmagnetized plasmas which
can be responsible for magnetic field generation if it becomes
unstable. In the derivation of its linear dispersion relation the
electron inertia was not ignored while the displacement current was
neglected. It was shown that the compressibility and vorticity can
couple due to density inhomogeneity and hence a low frequency
electromagnetic wave can be produced. Basically electrostatic IAW
and magnetostatic mode [2] cooperate with each other to develop such
a wave. The electron temperature perturbation was not taken into
account and the steady state was assumed to be maintained by
external mechanisms. The theory was applied to explain the magnetic
field generation in laser plasmas. But the longitudinal and
transverse characters of electric field decouple if the
quasi-neutrality is used. Then both the high frequency and low
frequency electromagnetic instabilities were also investigated in
unmagnetized plasmas [3]. \\The interest in the investigation of low
frequency magnetic fluctuations in unmagnetized plasmas was
initiated after the first experimental observation of magnetic field
generated in a laser induced plasma [4]. Then more experiments were
performed [5, 6] on these lines. Several theoretical models were
presented to explain the magnetic
field generation in laser produced plasmas [7-13].\\
The magnetic electron drift vortex (MEDV) mode was proposed as a
pure transverse linear mode which can exist because of the electron
temperature fluctuations in unmagnetized inhomogeneous electron
plasmas [13]. This mode can become unstable [14] if the equilibrium
electron temperature gradient is parallel to the density gradient
and it is
maintained by external effects.\\
Most of the mechanisms proposed to explain magnetic fluctuations in
initially unmagnetized plasmas are based on electron
magnetohydrodynamics (EMHD) which has been discussed in detail in
Refs. [15, 16].\\ Recently [17], EMHD has been used to study
long-lived and slowly propagating nonlinear whistler structures
(NLWS), or whistler spheromaks (WSPS) observed in laboratory
experiments [18, 19]. In MEDV mode the divergence of electric field
is assumed to be zero $(\nabla.\textbf{E}_1)=0$ while the divergence
of electron velocity is non-zero $(\nabla.\textbf{v}_{e1}\neq0)$.
Furthermore the electrons are inertia-less and ions are treated to
be stationary. The frequency $\omega$ of the MEDV mode is assumed to
lie in between the ion plasma frequency and electron plasma
frequency i.e. $\omega_{pi}<<\omega<<\omega_{pe}$ where
$\omega_{pj}=\left(\frac{4\pi n_{0j}e^{2}}{m_{j}}\right)$, for j=e,i
and c is speed of light, while k is the wave vector. These
restrictions and assumptions are indeed very strict. \\\\It is
important to find out some electromagnetic mode taking into account
the ion dynamics so that the strict restrictions on the frequency
and wavelength of the perturbation are relaxed. Then the only
required condition on frequency should be $\omega<<\omega_{pe}, ck$.
Furthermore, the electron temperature gradient is necessary for
hydrostatic steady state $\nabla p_{e0}=0$. In this case the
temperature gradient becomes anti-parallel to density. But laser and
astrophysical plasmas  are open systems and many external mechanisms
can maintain a study state with parallel density and temperature
gradients. For example, in stellar cores, both the density and
temperature increase towards centre of the star due to fusion and
star is held intact because of gravity. Therefore both the parallel
and anti-parallel gradients can be discussed. The electron thermal
fluctuations can produce electromagnetic wave as was proposed many
decades ago [13]. Several interesting works have appeared in
literature on these lines (see for example review articles [9, 15,
16]).\\\\ It is pointed out here that it is not necessary to assume
a pure transverse perturbation in an electron plasma. Rather the
perturbation can be partially transverse and partially longitudinal
and hence the electron density perturbation may not be neglected.
The frequency of such a wave in electron plasma turns out to be near
$v_{te}\kappa_{n}$ where
$v_{te}=\left(\frac{T_{e}}{m_{e}}\right))^{\frac{1}{2}}$ is the
electron thermal speed and
$\kappa_{n}=|\frac{1}{n_{0}}\frac{dn_{0}}{dx}|$ is the inverse of
density gradient scale length $L_{n}=\frac{1}{\kappa_{n}}$. In the
local approximation we need to have $\kappa_{n}<<k$. Furthermore the
condition $\omega_{pi}^{2}<<v^{2}_{te}k^{2}$ can be satisfied if
$\frac{m_{e}}{m_{i}}<<\lambda^{2}_{De}k^{2}_{y}$. Therefore,
generally we may have $v_{te}\kappa_{n}\lesssim\omega_{pi}$ and
hence one cannot neglect ion dynamics. Moreover,
$v_{te}\kappa_{n}\simeq c_{s}k$ (where
$c_{s}=\left(\frac{T_{e}}{m_{i}}\right)^{\frac{1}{2}}$ is ion sound
speed) is also possible. \\Therefore, it is expected that the
electron thermal fluctuations can couple with IAW to produce stable
and unstable low frequency electromagnetic waves in unmagnetized
plasmas. We are interested mainly in finding out a quadratic
dispersion relation for a low frequency electromagnetic wave to see
a simplest physical picture. For a comparison and general interest,
the previous theroretical models are also briefly discussed. \\In
the next section, the pure electron plasma is investigated including
the description of MEDV mode. In section III, the coupling of IAW
with magnetostatic mode due to inhomogeneity is briefly revised. In
section IV a new electromagnetic mode which arises due to the
coupling of IAW with thermal fluctuations is presented.In section V
the result is applied to a case of classical plasma for
illustration. Finally the physical assumptions and mathematical
limits of the results are discussed.

\textbf{II. Electron Plasma} \\Let us consider the electron plasma in the background of stationary ions.\\
The set of equations in the linear limit can be written as, $$m_e
n_0
\partial_t \textbf{v}_{e1}=-en_0 \textbf{E}_1-\nabla
p_{e1}\eqno{(1)}$$$$\nabla\times\textbf{B}_1=\frac{4\pi}{c}\textbf{J}_1\eqno{(2)}$$$$\textbf{J}_1=-en_0
\textbf{v}_{e1}\eqno{(3)}$$$$\nabla\times\textbf{E}_1=-\frac{1}{c}\partial_t
\textbf{B}_1\eqno{(4)}$$ Since
$$p_1=n_0 T_{e1}$$ therefore energy equation becomes,
$$\frac{3}{2}n_{0}\partial_{t}T_{e1}+\frac{3}{2}n_0
(\textbf{v}_{e1}.\nabla)T_{e0}=-p_{0}\nabla.\textbf{v}_{e1}\eqno(5)$$
First we describe the pure transverse MEDV mode. \\Curl of $(1)$
gives,
$$\partial_{t}(\nabla\times\textbf{v}_{e1})=\frac{e}{m_{e}c}\partial_{t}\textbf{B}_{1}
+\frac{1}{m_{e}n_{o}}(\nabla n_0)\times\nabla T_{e1}\eqno(6)$$
Equations $(2)$ and $(3)$ yield, $$\textbf{v}_{e1}=-\frac{c}{4\pi
en_{0}}\nabla\times\textbf{B}_{1}\eqno(7)$$ and hence
$$\nabla\times\textbf{v}_{e1}=\frac{c}{4\pi
en_{0}}\nabla^{2}\textbf{B}_{1}\eqno(8)$$ where $\nabla
n_{0}\times(\nabla\times\textbf{B}_{1})=0$ due to the assumption
$\textbf{k}\bot\nabla n_{0}\bot\textbf{B}_{1}.$ Equation (8)
predicts $\textbf{E}_{1}=E_{1}\mathbf{\hat{x}}$ while $\nabla
n_{0}={\bf{\hat{x}}}\frac{dn_{0}}{dx}$, $\nabla=(0,ik_y,0)$ and
$\textbf{B}_1=B_1 \hat{\textbf{z}}$ have been chosen. \\
Equation (6) and (8) yield,
$$(1+\lambda^{2}_{e}k^{2})\partial_{t}\textbf{B}_{1}=-\frac{c}{en_{0}}(\nabla
n_{0}\times\nabla T_{e1})\eqno(9)$$ where
$\lambda_{e}=\frac{c}{\omega}_{pe}$. Equation (7) gives,
$$\nabla.\textbf{v}_{e1}=\frac{c}{4\pi n_{0}e}\frac{\nabla
n_{0}}{n_{0}}.(\nabla\times\textbf{B}_{1})\eqno(10)$$ and therefore
one obtains,
$$T_{e1}=\frac{2}{3}\frac{c}{4\pi n_{0}e}k_{y}\kappa_{n}B_{1}\eqno(11)$$
Then (9) and (11) give the linear dispersion relation of MEDV mode
as,
$$\omega^{2}=\frac{2}{3}C_{0}(\frac{\kappa_{n}}{k_{y}})^{2}v^{2}_{Te}k_{y}^2\eqno(12)$$
where $C_{0}=\frac{\lambda^{2}_{e}k^{2}}{1+\lambda^{2}_{e}k^{2}}$
and $v_{te}=(T_e/m_e)^{\frac{1}{2}}$. If $\nabla T_{e0}\neq0$ is
assumed, then $(12)$ becomes,
$$\omega^{2}=L_{0}\frac{\kappa_{n}}{k_{y}}\left[\frac{(\frac{2}{3}\kappa_{n}
-\kappa_{T})}{k_{y}}\right]v^{2}_{Te}k^{2}\eqno(13)$$
where $\kappa_{T}=|\frac{1}{T_{e0}}\frac{dT_{e0}}{dx}|$ and $\nabla
T_{e0}=+\textbf{{x}}\frac{dT_{e0}}{dx}$ has been used. If
$(\frac{2}{3}\kappa_{n}-\kappa_{T})<0,$ then the mode becomes
unstable[14].\\It can be noticed that we have
$\nabla.\textbf{E}_1=0$ and $\nabla.\textbf{v}_{e1}\neq0$ in the
above treatment and it does not seem to be very convincing.\\ If we
write (1) in x and y components, we can observe that the y-component
of $\textbf{E}_1$ should not be considered as zero due to
$\partial_y p_{e1}\neq0$ and hence the wave may be partially
transverse and partially longitudinal.\\ The curl of (2) yields a
relation between $\textbf{E}_{1x}$ and $\textbf{E}_{1y}$,
$$\textbf{E}_{1x}=-\frac{1}{a}\frac{\kappa_n}{k_y}(iE_{1y})\eqno{(14)}$$
where $a=(1+\lambda_{e}^{2} k_{y}^{2})$.\\ Equation (5) becomes,
$$W_{0}^{2}\frac{T_{e1}}{T_0}=v_{te}^{2} k_{y}^{2}\left(1-\frac{3}{2}\frac{\kappa_{T}^{2}}{k_{y}^{2}}
\right)\frac{n_{e1}}{n_0}-\frac{e}{m_e}\left(ik_y E_{1y}+\frac{3}{2}k_T
E_{1x}\right)\eqno{(15)}$$ where $W_{0}^{2}=\frac{3}{2}\omega^2
-v_{te}^{2} k_{y}^{2}\left(1-\frac{3}{2}\frac{\kappa_{T}
\kappa_n}{k_{y}^{2}}\right)$ \\The continuity equation yields,
$$L_{0}^{2}\frac{n_{e1}}{n_0}=-\left(1+\frac{v_{te}^{2} k_{y}^{2}}{W_{0}^{2}}\right)\left(i\frac{e}{m_e}k_y
E_{1y}\right)$$$$-\frac{e}{m_e}\left(1+\frac{3}{2}\frac{\kappa_T}{\kappa_n}\frac{v_{te}^{4}
k_{y}^{4}}{W_{0}^{2}}\right)\kappa_n E_{1x}\eqno{(16)}$$ where
$L_{0}^{2}=\left\{\omega^2-v_{te}^{2}k_{y}^{2}-\frac{v_{te}^{4}k_{y}^{2}}{W_{0}^{2}}\left(1-\frac{3}{2}
\frac{\kappa_{T}^{2}}{k_{y}^{2}}\right)\right\}$. The Poisson
equation
$$\nabla.\textbf{E}_1=-4\pi
e\left(\frac{n_{e1}}{n_0}\right)\eqno{(17)}$$ can be written as,
$$\left\{\left[\frac{3}{2}\omega^4 -v_{te}^{2}k_{y}^{2}\left(\frac{3}{2}+G_n\right)\omega^2 -\frac{3}{2}
v_{te}^{4}k_{y}^{4}(g_{T}^{2}-g_{nT}^{2})\right]-\omega_{pe}^{2}\left[\frac{3}{2}\omega^2
+\frac{3}{2}v_{te}^{2}k_{y}^{2}g_{nT}^{2}\right]\right\}iE_{1y}$$
$$=\omega_{pe}^{2}\left[\frac{3}{2}\omega^2
-v_{te}^{2}k_{y}^{2}\left(1-\frac{3}{2}g_{nT}^{2}-\frac{3}{2}\frac{\kappa_T}{\kappa_{n}}\right)\right]
\frac{\kappa_n}{k_y}iE_{1x}\eqno{(18)}$$
where $G_n=\left(1-\frac{3}{2}g_{nT}^{2}\right)$,
$g_{nT}^{2}=\frac{\kappa_n \kappa_T}{k_{y}^{2}}$ and
$g_{T}^{2}=\kappa_{T}^{2}/k_{y}^{2}$. \\Equations (14) and (18)
yield a linear dispersion relation in the limit $\omega^2 <<
\omega_{pe}^{2}$ as, $$\omega^2 \simeq
\frac{2}{3H_0}\left[-v_{te}^{2}k_{y}^{2}\left(\frac{\kappa_{n}^{2}}{k_{y}^{2}}+\frac{3}{2}
\lambda_{e}^{2}k_{y}^{2}g_{nT}^{2}\right)+\frac{3}{2}a\lambda_{De}^{2}k_{y}^{2}(g_{T}^{2}-g_{nT}^{2})
\right]\eqno{(19)}$$
where
$H_0=\left\{(1+\lambda_{De}^{2}k_{y}^{2})a-\frac{\kappa_{n}^{2}}{k_{y}^{2}}+a\lambda_{De}^{2}k_{y}^{2}G_n\right\}$
and $\lambda_{De}^{2}=\left(\frac{T_e}{4\pi n_0 e^2}\right)$. The
last term on right hand side of (19) is smaller because
$\lambda_{De}^{2}k_{y}^{2}<1$ and due to local approximation we have
$\kappa_n, \kappa_T << k_y$. Therefore (19) can be simplified as,
$$\omega^2=-\frac{2}{3H_0}v_{te}^{2}\kappa_{n}^{2}\left(1+\frac{3}{2}\lambda_{e}^{2}k_{y}^{2}
\frac{\kappa_T}{\kappa_n}\right)\eqno{(20)}$$ which predicts a
purely growing electromagnetic mode.\\ The dispersion relation (13)
can not be retrieved from (20) because we have evaluation
$\nabla.\textbf{v}_{1e}$ from equation of motion and in the
derivation of (13), the velocity $\textbf{v}_{e1}$ is evaluated
using Ampere's law for pure transverse wave. The frequency $\omega$
in both the dispersion relations is of the order of $v_{te}^{2}
\kappa_{n}^{2}$ or smaller.\\ The important point to note is that
$$\omega^2 \sim \left(\frac{\kappa_n}{k_y}\right)^2 (v_{te}^{2}
k_{y}^{2})=\left(\frac{\kappa_n}{k_y}\right)^2
(\lambda_{De}^{2}k_{y}^{2})\omega_{pe}^{2}\eqno{(21)}$$ and
$\lambda_{De}^{2}k_{y}^{2}<<1$ has been assumed to ignore electron
plasma wave. Therefore, even for
$\frac{m_e}{m_i}<\left(\frac{\kappa_n}{k_y}\right)^2$, the frequency
in (21) can satisfy the relation $\omega \lesssim \omega_{pi}$ or
even $\omega$ can be of the order of IAW frequency $c_s k_y$.\\
Therefore, the partially transverse and partially longitudinal
electron wave of equation
(20) may couple with ion acoustic mode.\\

\textbf{III. IAW and Magnetostatic Mode} \\Here we shall show that
transverse magnetostatic mode [2] which is obtained in the limit
$\omega^2<<\omega_{pe}^{2}$ can couple with IAW in a nonuniform
plasma [1]. Therefore, both ions and electrons are considered to be
dynamic. The ions are assumed to be cold for simplicity and
therefore equation of motion becomes,
$$\partial_{t}\textbf{v}_{i1}=\frac{e}{m_{i}}\textbf{E}_{1}\eqno(22)$$
The continuity equation yield,
$$\frac{n_{i1}}{n_{0}}=\frac{e}{m_{i}\omega^{2}}(\kappa_{n}E_{1x}+ik_{y}E_{1y})\eqno(23)$$
If quasi-neutrality is used then transverse component $E_{1x}$ and
longitudinal component $E_{1y}$ become uncoupled. Therefore, we
assume $\lambda^{2}_{De}k^{2}_{y}\neq0$ and use Poisson equation
which in the limit $\omega^{2}<<\omega^{2}_{pe}$ becomes,
$$\left[-v^{2}_{te}k^{2}_{y}\omega^{2}-\omega^{2}_{pi}(\omega^{2}-v^{2}_{te}k^{2}_{y})-\omega^{2}_{pe}\omega^{2}\right]ik_{y}E_{1y}\simeq[\omega^{2}_{pi}(\omega^{2}-v^{2}_{te}k^{2}_{y})+\omega^{2}_{pe}\omega^{2}]\kappa_{n}E_{1x}\eqno(24)$$
Note that the term $v^{2}_{te}k^{2}_{y}$ is not ignored compared to
$\omega^{2}_{pe}$ to couple $E_{1x}$ and $E_{1y}$.\\ Equations (14)
and (24) give a linear dispersion relation as,
$$\omega^{2}=\frac{c^{2}_{s}k^{2}_{y}(a-\frac{\kappa^{2}_{n}}{k^{2}_{y}})}{(ab-\frac{\kappa^{2}_{n}}{k^{2}_{y}})}\eqno(25)$$
where $c^{2}_{s}=\frac{T_{e}}{m_{i}}$, and
$b=(1+\lambda^{2}_{De}k^{2}_{y})$. We have to use Poisson equation
to obtain a quadratic equation in $\omega$ while (2) implies
$n_{e1}\simeq n_{i1}$. The equation (25) is the same as equation
(21) of Ref. [1] where two small terms in the denominator are
missing. Note that $\lambda^{2}_{De}<\lambda^{2}_{e}$ and if quasi
neutrality is used due to Ampere's law, then (25) yields the basic
electrostatic IAW dispersion relation
$\omega^{2}=c^{2}_{s}k^{2}_{y}$. If displacement current is retained
and Poisson equation is used without using $\omega^{2},
\omega^{2}_{pi}<<\omega^{2}_{pe}$, then one obtains a dispersion
relation of coupled three waves; ion acoustic wave, electron plasma
wave and high frequency transverse wave [3].
\\Actually the contribution of displacement current in the curl of
Maxwell's equation has been neglected for
$\omega^{2}<<\omega^{2}_{pe}, c^{2}k^{2}$. This should not mean that
the electrostatic part of current is also divergence free, in our
opinion. In the divergence part of Maxwell's equation
$\omega^{2}<<\omega^{2}_{pe}$ is used but $v^{2}_{te}k^{2}_{y}$ term
is assumed to be important. It may be mentioned that in this
treatment, Ampere's law does not imply quasi-neutrality necessarily.
We need a coupling of divergence part and cure part of the current
and for this we need to assume $1<\lambda_{De}^{2}k_{y}^{2}$ in the
limit
$1<\omega^2<<\omega_{pe}^{2}, c^2 k^2$.\\\\

\textbf{IV. Unstable Electromagnetic Wave} \\ Now we present a
simple but interesting picture of ideal plasma assuming ions to be
cold. The divergence and curl of (1) give, respectively,
$$\partial_{t}\nabla.(n_{0}\textbf{v}_{e1})=-\frac{e}{m_{e}}n_{0}\nabla.\textbf{E}_{1}-\frac{e}{m_{e}}\nabla
n_{0}.\textbf{E}_{1}-\frac{1}{m_{e}}(\nabla.\nabla
p_{e1})\eqno(26)$$ and
$$\partial_{t}(\nabla\times\textbf{v}_{e1})+(\mathbf{\kappa}_{n}\times\partial_{t}\textbf{v}_{e1})=
-\frac{e}{m_{e}}\mathbf{\kappa}_{n}\times\textbf{E}_{1}-\frac{e}{m_{e}}\nabla\times\textbf{E}_{1}\eqno{(27)}$$
where $\kappa_{n}=|\frac{1}{n_{0}}\frac{dn_{0}}{dx}|$ and $\nabla
n_{0}=+\mathbf{\hat{x}}|\frac{dn_{0}}{dx}|$ has been assumed. If
initially electric field was purely electrostatic i.e.
$\textbf{E}_{1}=-\nabla\varphi_{1}$, then it will develop a rotating
part as well if $\nabla n_{0}\times\textbf{E}_{1}\neq0$, as is
indicated by the right hand side (RHS) of (27). \\Therefore, in an
inhomogeneous plasma electron thermal fluctuations can take place
due to low frequency perturbations of electric field which has both
transverse and longitudinal components. Then these thermal
fluctuations can couple with IAW even in the quasi-neutrality
limit.\\ However, we use the Poisson equation
$$\nabla.\textbf{E}_1=4\pi n_0 e \left(\frac{n_{i1}}{n_0}-\frac{n_{e1}}{n_0}\right)\eqno{(28)}$$ Using (16) and (23),
the above equation can be written as,
$$
[(L^{2}_{0}W^{2}_{0})\omega^{2}-(L^{2}_{0}\omega^{2}_{0})\omega^{2}_{pi}-\omega^{2}_{pe}(W^{2}_{0}+v^{2}_{te}k^{2}_{y})]
ik_{y}E_{1y}=
$$
$$
\left[L^{2}_{0}W^{2}_{0}\omega^{2}_{pi}+\omega^{2}_{pe}\left(W^{2}_{0}+
\frac{3}{2}\frac{\kappa_{T}}{\kappa_{n}}v^{2}_{te}k^{2}_{y}\right)\omega^{2}\right]\kappa_{n}E_{1x}\eqno{(29)}
$$
In the limit $\omega^2, \omega_{pi}^{2}<<\omega_{pe}^{2}$, (29)
reduces to,

$$
[\left\{ -\frac{3}{2}\omega _{pe}^{2}-v_{te}^{2}k_{y}^{2}\left(
\frac{3}{2} +G_{n}\right) \right\} \omega ^{4}$$ $$
 +\left\{
\frac{3}{2}v_{te}^{4}k_{y}^{4}(g_{T}^{2}-g_{nT}^{2})-\frac{3}{2}
\omega _{pe}^{2}v_{te}^{2}k_{y}^{2}g_{nT}^{2}+\omega
_{pi}^{2}v_{te}^{2}k_{y}^{2}\left( \frac{3}{2}+G_{n}\right)
\right\}\omega^2 $$ $$
 -\frac{3}{2}\omega
_{pi}^{2}v_{te}^{4}k_{y}^{4}(g_{T}^{2}-g_{nT}^{2})] iE_{1y}
$$ $$
 =[\frac{3}{2}\omega _{pe}^{2}\omega ^{4}-\left\{ \omega
_{pi}^{2}v_{te}^{2}k_{y}^{2}\left(\frac{3}{2}+G_{n}\right)-\frac{3}{2}\omega
_{pe}^{2}v_{te}^{2}k_{y}^{2}\frac{\kappa _{T}}{\kappa
_{n}}+\omega_{pe}^{2}v_{te}^{2}k_{y}^{2}G_n\right\} \omega^{2}
$$ $$
+\frac{3}{2}\omega
_{pi}^{2}v_{te}^{4}k_{y}^{4}(g_{T}^{2}-g_{nT}^{2})]\left(
\frac{\kappa _{n}}{k_{y}}E_{1x}\right)\eqno{(30)}
$$

Equations (14) and (30) yield a linear dispersion relation as
follows,
$$
\frac{3}{2}\left[ ab-\left( \frac{\kappa _{n}}{\kappa
_{y}}\right)^{2}+\frac{2}{3}a\lambda _{De}^{2}k_{y}^{2}G_{n}\right]
\omega ^{4}
$$ $$
+[-\frac{3}{2}\lambda
_{De}^{2}k_{y}^{2}v_{te}^{2}k_{y}^{2}(g_{T}^{2}-g_{nT}^{2})a+\frac{3}{2}%
v_{te}^{2}k_{y}^{2}g_{nT}^{2}a-c_{s}^{2}k_{y}^{2}\left( \frac{3}{2}%
+G_{n}\right)a
$$ $$
+\left( \frac{\kappa _{n}}{k_{y}}\right)^{2}\left\{
c_{s}^{2}k_{y}^{2}\left( \frac{3}{2}+G_{n}\right) +v_{te}^{2}k_{y}^{2}G_{n}-%
\frac{3}{2}v_{te}^{2}k_{y}^{2}\frac{\kappa _{T}}{\kappa
_{n}}\right\}]\omega^2
$$ $$
+\frac{3}{2}%
ac_{s}^{2}k_{y}^{2}v_{te}^{2}k_{y}^{2}(g_{T}^{2}-g_{nT}^{2})-\left( \frac{%
\kappa _{n}}{k_{y}}\right)
^{2}c_{s}^{2}k_{y}^{2}v_{te}^{2}k_{y}^{2}(g_{T}^{2}-g_{nT}^{2})=0%
\eqno{(31)}
$$%
This equation contains a coupling of both the electromagnetic
waves described by the dispersion relations (20) and (25). We are
interested in finding out a simple dispersion relation in quadratic
form to look at the physical phenomena more clearly.\\ For this we
assume $g_{nT}=g_T$ i.e. $\kappa_T=\kappa_n$ and in this case (31)
becomes,
$$\omega^2=\frac{2}{3H_0}\left\{c_{s}^{2}k_{y}^{2}\left(\frac{3}{2}+G_n\right)
\left(a-\kappa_{n}^{2}/k_{y}^{2}\right)-v_{te}^{2}\kappa_{n}^{2}
\left(1+\frac{3}{2}\lambda_{e}^{2}k_{y}^{2}\right)\right\}\eqno{(32)}$$
where
$H_0=\left[ab-\frac{\kappa_{n}^{2}}{k_{y}^{2}}+\frac{3}{2}a\lambda_{De}^{2}k_{y}^{2}G_n\right]$.
In the quasi-neutrality limit, we have b=1 in the above
electromagnetic dispersion relation and it reduces to
$$\omega^2\simeq
\frac{2}{3H_1}\left\{c_{s}^{2}k_{y}^{2}\left(\frac{5}{2}a
\frac{\kappa_{n}^{2}}{k_{y}^{2}}\left(a+\frac{5}{2}\right)\right)
-v_{te}^{2}\kappa_{n}^{2}\left(1+\frac{3}{2}\lambda_{e}^{2}k_{y}^{2}\right)\right\}\eqno{(33)}$$
where $H_1=\left(a-\kappa_{n}^{2}/k_{y}^{2}\right)$ and for
$\kappa_n=0$, we obtain $\omega^2=\frac{5}{3}c_{s}^{2}k_{y}^{2}$.\\
It is important to mention that the transverse and longitudinal
components of electric field do not decouple in the presence of
electron thermal fluctuations even if quasi-neutrality is used. If
thermal fluctuations are ignored (32) reduces to (25) and if ion
dynamics is neglected we recover the result of equation (20). In
principle both the waves described by (20) and (25) have frequencies
very near to each other under local approximation in plasmas of
low-Z materials. Therefore the coupled mode (32) is a fundamental
unstable electromagnetic wave of unmagnetized inhomogeneous plasmas.
If $\lambda_{e}^{2}k_{y}^{2}<m_e/m_i$, then $a\simeq1$ and b=1. In
this case Eq. (33) predicts purely growing electrostatic
fluctuations for
$\frac{m_e}{m_i}<\left(\frac{\kappa_n}{k_y}\right)^2$ because the
second term on RHS of (33) can become larger than the first term.\\\\\\
\textbf{V. Application}\\
As an illustration we consider the parameters of classical
laser-induced plasma as $n_0\sim10^{20}cm^{-3}$ and $T_e=100ev$.
Then we obtain $c_s \sim 10^7 cm/Sec$, $v_{te}=4.18\times 10^{8}
cm/Sec$, $\lambda_e \sim 0.53 \times 10^{-4}cm$, $\lambda_{De}\sim
0.74 \times 10^{-6} cm$, and $\omega_{pi}\sim 1.3 \times 10^{13}
rad/Sec$. If $\kappa_n \sim 10^4 cm^{-1}$ and we choose $k_y \sim
10^5 cm^{-1}$, in Hydrogen plasma we find
$\frac{m_e}{m_i}<\left(\frac{\kappa_n}{k_y}\right)^2 =10^{-2}$. We
obtain a=29, b=1.0055, $v_{te}\kappa_n \sim 4.18 \times 10^{12}
rad/Sec$ and $c_s k_y \sim 10^{12} rad/Sec$. Since
$\frac{m_e}{m_i}<\lambda_{De}^{2}k_{y}^{2}$, we use the result of
Poisson equation (32) and obtain $$\omega\simeq \pm i(5.8 \times
10^{12})\eqno{(34)}$$ Note that it gives a purely growing
electromagnetic instability. Furthermore $c_s
k_y\lesssim|\omega|\lesssim\omega_{pi}$ and hence
one should not ignore ion dynamics.\\
The instability appears due to the second term on RHS which enters
because of pure electron dynamics. But fluctuations are at ion time
scales. Therefore, ions must be treated to be mobile.
The instability can also appear in the quasi-neutrality limit.\\
If $\kappa_n \sim 10^3 cm^{-1}$ and we choose $k_y \sim 10^4
cm^{-1}$, then $\lambda_{De}^{2} k_{y}^{2}<m_e/m_i$ and hence we use
quasi-neutrality. Then a=1.28 and (33) yields, $$\omega\simeq\pm i
(3.6\times 10^{12})\eqno{(35)}$$ Again we see a purely growing
electromagnetic mode
with $|\omega|\lesssim \omega_{pi}$.\\\\
\textbf{VI. Discussion}\\
A low frequency unstable electromagnetic wave has been studied in
unmagnetized inhomogeneous plasmas. This wave can be an intrinsic
source of magnetic field fluctuations in initially unmagnetized
plasmas. Dynamics of both electrons and ions are important in this instability.\\
These fluctuations can become unstable when the density and
temperature gradients are parallel to each other as is the case, in
the stellar cores. Similarly in laser-induced plasmas these
gradients can be in the
same direction due to external conditions.\\
In a statistical ensemble of charged particles, the electrostatic
fluctuations are quite natural. Any initial electrostatic field
perturbation can produce its transverse component as well in the
presence of density gradient. This phenomenon can cause a coupling
of ion acoustic wave (IAW) with the low frequency transverse
magnetostatic mode. This coupled mode has already been investigated
more than a decade ago [1]. But it can exist in a relatively shorter
wavelength range for $\frac{m_e}{m_i}<\lambda_{De}^{2}k^2$. In the
quasi-neutrality limit, the IAW and magnetostatic modes decouple.\\
Another interesting mechanism has been proposed for the generation
of a transverse wave in as electron plasma. In this case the
transverse wave (MEDV mode) is produced by the electron temperature
perturbation [13, 14]. This mode does not have density fluctuations.
In the present investigations it has been shown that a low frequency
electromagnetic wave seems to appear in a non-uniform unmagnetized
electron plasma having both the longitudinal and transverse electric
field components. But the frequencies of both the above mentioned
modes are of the order of $v_{te}\kappa_n$ or even smaller.\\ Since
frequencies $v_{te}\kappa_n$ and $c_{s}k_y$ can be very near to each
other, therefore ions should not be assumed as stationary. We need
to consider ion dynamics as well.Then we obtain a coupled linear
dispersion relation which is fourth order polynomial in $\omega$. If
$\kappa_T=\kappa_n$ is assumed, then it reduces to a quadratic
equation (32). It indicates the presence of a partially transverse
and partially longitudinal unstable wave in nonuniform unmagnetized
electron-ion plasmas. Dynamics of both electrons and ions take part
in the generation of this instability. The electron temperature
fluctuations are necessary to produce this wave. It is also
important to mention that this wave can exist in the
quasi-neutrality limit as well.\\ In our opinion, this
electromagnetic instability plays an important role in the
generation of fluctuating magnetic fields in initially unmagnetized
plasmas. The nonlinear saturation mechanisms need to be investigated
which can contribute to the magnetization of the plasmas. Therefore
the electromagnetic waves and instability presented here can be
important in the study of magnetic field generation in laser-induced
and astrophysical plasmas.
\pagebreak

\end{document}